\documentclass{webofc}
\usepackage[varg]{txfonts}   

\usepackage{footmisc}

\usepackage{enumerate}

\def\mkfit{mkFit\xspace}

\def\stt#1{{\small\texttt{#1}}}

\def\etal{\emph{et al.}\xspace}

\def\MeVoc{\ensuremath{\,\rm{M}e\rm{V}/c}}

\def\GeVoc{\ensuremath{\,\rm{G}e\rm{V}/c}}

\def\twop{0.48\textwidth}

\def\postfigskip{\vskip-4mm}

\begin{document}

\title{Parallelized and Vectorized Tracking Using Kalman Filters with CMS
  Detector Geometry and Events}

\author{
       \firstname{Giuseppe} \lastname{Cerati}\inst{4} 
  \and \firstname{Peter} \lastname{Elmer}\inst{2} 
  \and \firstname{Brian} \lastname{Gravelle}\inst{5} 
  \and \firstname{Matti} \lastname{Kortelainen}\inst{4} 
  \and \firstname{Vyacheslav} \lastname{Krutelyov}\inst{1} 
  \and \firstname{Steven} \lastname{Lantz}\inst{3} 
  \and \firstname{Matthieu} \lastname{Lefebvre}\inst{2} 
  \and \firstname{Mario} \lastname{Masciovecchio}\inst{1} 
  \and \firstname{Kevin} \lastname{McDermott}\inst{3} 
  \and \firstname{Boyana} \lastname{Norris}\inst{5} 
  \and \firstname{Allison} \lastname{Reinsvold Hall}\inst{4} 
  \and \firstname{Daniel} \lastname{Riley}\inst{3} 
  \and \firstname{Matev\v{z}} \lastname{Tadel}\inst{1}\fnsep\thanks{\email{mtadel@ucsd.edu}}	
  \and \firstname{Peter} \lastname{Wittich}\inst{3} 
  \and \firstname{Frank} \lastname{W\"{u}rthwein}\inst{1} 
  \and \firstname{Avi} \lastname{Yagil}\inst{1} 
}

\institute{UC San Diego, La Jolla, CA, USA 92093
  \and     Princeton University, Princeton, NJ, USA 08544
  \and     Cornell University, Ithaca, NY, USA 14853
  \and     Fermilab, Batavia, IL, USA 60510-5011
  \and     University of Oregon, Eugene, OR, USA 97403
}

\abstract{%
  The High-Luminosity Large Hadron Collider at CERN will be characterized by
  greater pileup of events and higher occupancy, making the track
  reconstruction even more computationally demanding. Existing algorithms at
  the LHC are based on Kalman filter techniques with proven excellent physics
  performance under a variety of conditions. Starting in 2014, we have been
  developing Kalman-filter-based methods for track finding and fitting adapted
  for many-core SIMD processors that are becoming dominant in high-performance
  systems.

  This paper summarizes the latest extensions to our software that allow it to
  run on the realistic CMS-2017 tracker geometry using CMSSW-generated events,
  including pileup. The reconstructed tracks can be validated against either
  the CMSSW simulation that generated the detector hits, or the CMSSW reconstruction of
  the tracks. In general, the code's computational performance has continued
  to improve while the above capabilities were being added. We demonstrate
  that the present Kalman filter implementation is able to reconstruct events
  with comparable physics performance to CMSSW, while providing generally
  better computational performance. Further plans for advancing the software
  are discussed.
}

\maketitle

\section{Introduction}

Over the past few years, plans for the High-Luminosity Large Hadron Collider
upgrade project, and the accompanying tenfold leap in luminosity, have made it
clear that a significant research and development effort is required towards the
2020 to 2025 timeframe to meet the increased complexity and computational
requirements of the track finding algorithms. The expected increase in event
complexity, coupled with the technological changes that continue to drive
interest in multi/many-core processors, have motivated the community to explore
radically different algorithms and computing architectures to address the
anticipated issues~\cite{CTD2018}. Our approach, however, has been to focus on
the traditional, well-known, and well-understood Kalman Filter (KF) method~\cite{Fruhwirth}, to
see how far KF-based tracking can be pushed in this new environment. To that
end we have been developing the \mkfit framework, designed from the ground up for
performance, that is better suited to utilize the types of parallelism available
in contemporary general-purpose computing hardware. The ultimate goal of the project is
to reach physics performance on par with the standard Compact Muon Solenoid (CMS)
tracking~\cite{CMS-tracking} while achieving a significant reduction in processing time.

All things considered, a fully-integrated, parallelized and vectorized implementation 
of a KF-based tracking application provides a strong reference point for the evaluation 
of more exotic types of solutions, both in terms of computational as well as physics performance. 
Indeed, new solutions of any type will have to demonstrate superior performance in both of these 
respects to operate successfully in environments with moderate time constraints such as high-level 
trigger applications and offline reconstruction.

The project was started in 2014 with the development of a
``Matriplex'' matrix operation library optimized for simultaneous vectorized
processing of sets of small matrices. From this basis, the initial
implementation of vectorized KF fitting was demonstrated on a simplified
barrel-only detector~\cite{pkf-fit}.

The next stage was the implementation of track finding using the above
technology and the same simplified geometry~\cite{pkf-finding}. Simple
multithreading was implemented by partitioning tracks in up to 21
$\eta$ bins and using OpenMP parallel constructs. Physics performance was
adequate (95\% efficiency and correct $\chi^2$ distribution of tracks and
pulls of the track parameters), but the achieved parallelization speedups were
a bit disappointing (x2 for vectorization and x10 for multithreading on Intel
KNC), indicating the need to decrease the fraction of non-vectorizable code
and implement a better work partitioning scheme. To this end, processing of
track candidates on each layer was optimized to reduce the number of
instantiations of Track objects by selecting only the hits giving the best
extensions to tracks based on their
$\chi^2$ score before doing the final Kalman updates~\cite{pkf-clone-engine}.

OpenMP was replaced by Intel Thread Building Blocks (TBB) to increase
flexibility as well as to be in compliance with the CMS code
base~\cite{pkf-tbb}. Further, to avoid imbalances in $\eta$ regions and to
provide more workload tasks for the many available cores, support for
processing of multiple events in parallel was added. This allowed the
individual tasks to remain relatively large while still being able to fill up
all available hardware threads~\cite{pkf-acat-17}.

Significant effort has been put into porting of \mkfit to run on GPGPUs using
CUDA~\cite{pkf-gpu}. Fitting and track finding have been ported for the
barrel-only simplified detector. Performance results for track finding were
disappointing with \mkfit only being able to use about 4\% of the available
GPU processing power. Nevertheless, Matriplex is observed to outperform
standard small-matrix multiplication packages for GPUs. We are currently in
the process of quantifying performance plateaus reachable for KF-like
operations as a function of problem size, problem segmentation, and arithmetic
intensity with the intention of identifying architectural limitations to
running KF-based track finding on GPGPUs.

Beginnning in 2015, \mkfit was extended incrementally to handle realistic
detector geometries with barrel and endcap sections. This required
implementation of the KF and propagation equations for the endcap case, as
well as a consolidated steering code that was able to handle both barrel and
endcap cases. Finally, a general detector description mechanism was
implemented to support arbitrary detector geometries.

Currently, \mkfit is able to run on CMS-2017 geometry with reasonable physics
and computational performance. Ongoing work is focusing on improving the
physics performance through fine-tuning of hit and track selection
algorithms. Post-processing of found tracks and duplicate track removal
still needs to be implemented or may be delegated to algorithms in CMS
software.

This paper focuses on recent developments required to fully support realistic
geometry of the CMS detector and to process simulated
CMS data with up to 70 minimum bias $pp$ collisions superimposed over the
signal $t\bar{t}$ events. Generalized geometry handling is described in
section \ref{sec:cms-geom-and-events} and physics and computational
performance are presented in sections \ref{sec:phys-perf} and
\ref{sec:comp-perf}.

\section{Handling of CMS geometry and events}
\label{sec:cms-geom-and-events}

\subsection{Geometry and Detector description}

Geometry in \mkfit is described as a vector of \stt{LayerInfo} structures
that contain the physical dimensions of a layer, hit search windows, and parameters
and flags relevant for track finding. This includes information about layer
detector type, stereo/mono layers, and an optional hole in detector
coverage as needed for the CMS endcap detectors (this could be extended for
even more general acceptance handling).

For track finding, \emph{steering parameters} need to be defined for every
\emph{tracking region}. So far, it has been sufficient to consider only five distinct $\eta$
regions (barrel, $+z$/$-z$ transition, and $+z$/$-z$ endcap) but the concept
could be used also to separate regions by $p_T$ or by tracking iteration. The
steering parameters contain, most importantly, a vector of \stt{LayerControl}
structures that hold the layer indices (mapping into the \stt{LayerInfo} vector)
that need to be traversed during track finding. Additionally, it contains
layer parameters and flags that are specific for this tracking region, such as
tags for layers that are possible seeding layers or layers to be considered only
during backward fitting. This allows the track finding algorithm to be completely
agnostic of the detector structure: it simply follows the layer propagation
plan in the steering parameters and executes operations in accordance with the
control flags in \stt{LayerControl} and \stt{LayerInfo} structures.

Geometry and steering parameter setup is implemented as a plugin that
populates the in-memory data structures with the required information. With
this functionality, we are able to support both a simple geometry used for development and
CMS-2017 geometry with all detector-specific information existing only in the
plugin code. For the CMS-2017 geometry, we include the effects of multiple
scattering and energy loss by defining two-dimensional arrays for the
radiation and interaction lengths that are quickly indexed in
$r-z$. These constants are taken from CMS simulation for the amount of
material a particle would traverse propagating from module to module. \mkfit
supports usage of both constant and parameterized magnetic field; either type of field can
be selected each time propagation is required in the code.

\subsection{Handling and processing of CMS events}
\label{ssec:cms-event-processing}

When processing CMS events \mkfit relies on hit and seed data to be provided
externally. In the standalone case (where mkFit operates independently of CMSSW),
\mkfit reads these data from a binary file
created by a converter application. Additionally, the binary file can also 
contain vectors of simulated tracks and reconstructed tracks as found by 
standard CMS tracking, for later use in the validation of \mkfit's performance.

Before passing seeds to \mkfit for track finding, the seed collection is
``cleaned'' by removing multiple instances of seeds that are most likely 
based on hits belonging to the same outgoing particle. The cleaning
algorithm uses the identity of hits and fitted seed parameters $p_T$, $\eta$,
and $\varphi$ to eliminate duplicate seeds and is tuned so as to not cause any
drop in track finding efficiency for high pile-up events. The duplicate seeds
arise due to detector module overlaps that are rather significant, especially
in the endcaps, where the modules on the same blade of a disk overlap considerably.

In principle, seed cleaning could be performed as a final step in the seed
finding algorithm; however, due to the way standard track finding works in
CMSSW, this was not deemed necessary. CMSSW processes seeds one by one and
when a track candidate is found, its hits are tagged as used. A seed is
rejected if all its hits have already been used by a track candidate found
earlier. The technique relies on CMSSW backward propagation to find additional
hits on the seeding layers, and reduces the duplicate seeds in the first step
of their consideration to a negligible level. This is
not possible in \mkfit where we process up to $32 \times N_{threads}$ seeds in parallel
and, as we try to group seeds that are close in $\eta$ and $\varphi$ to
maximize memory cache reuse of hit data, this could lead to significant waste
of processing slots.

As already mentioned, we have recently started the process of including \mkfit
in standard CMS software distribution. \mkfit is used as an external software
package with a dedicated CMS processing module running within the CMS
framework. This module packages the input data (seeds and hits) in format
expected by \mkfit, and provides high-level configuration and steering of
\mkfit execution. When an event is processed, it copies resulting tracks back
into CMS format. This mode of inclusion allows \mkfit code to remain
independent of CMS particularities and overhead as well as allows us to
perform development and testing in a more lightweight environment.

\section{Physics performance}
\label{sec:phys-perf}

This section presents current basic physics performance plots for \mkfit
running on CMS-2017 geometry with a CMSSW simulated sample of 5000 $t\bar{t}$
events, each of which has been superimposed with a mean of 70 minimum-bias $pp$ collisions.
Constant magnetic field of 3.8\,T has been used. We are showing
results corresponding to the CMS \emph{initial tracking iteration} where seeds are
required to have 4 hits all coming from distinct inner pixel layers and be
compatible with the beam spot constraint. We show equivalent results from CMSSW using
the same set of input seeds.

While these results show the actual performance of \mkfit, they are
preliminary in the sense that we know further work is necessary to make a fair
comparison between CMSSW and \mkfit initial iteration tracking:
\begin{enumerate}[--]\topsep-2pt\itemsep-2pt
\item \mkfit's hit selection windows, candidate scoring criteria, and final track
  quality criteria have not yet been tuned for optimal performance.
\item Cleaning and merging of the final track collection have not yet been
  implemented in \mkfit. This includes removal of duplicate tracks due to
  multiple seeds per particle.
\item To ensure a fair comparison of efficiency, the same final track selection
  criteria and post-processing need to be applied for both algorithms. CMSSW
  intentionally uses stronger requirements in initial iteration, relying on later
  iterations to pick up less likely track candidates.
\end{enumerate}

\begin{figure}[thb]
  \centering
  \includegraphics[width=\twop]{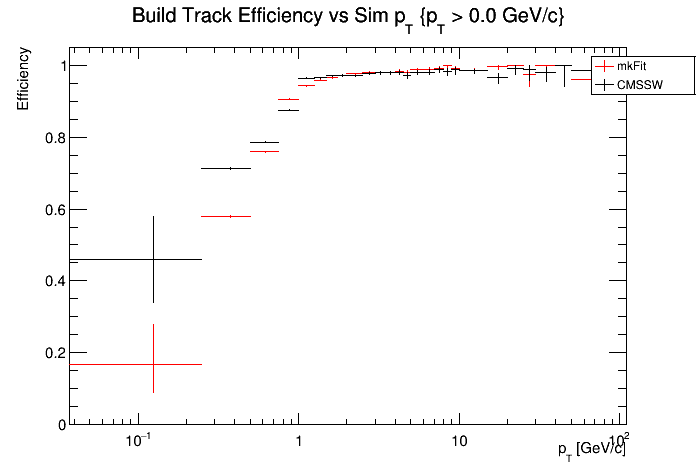}
  \hfill
  \includegraphics[width=\twop]{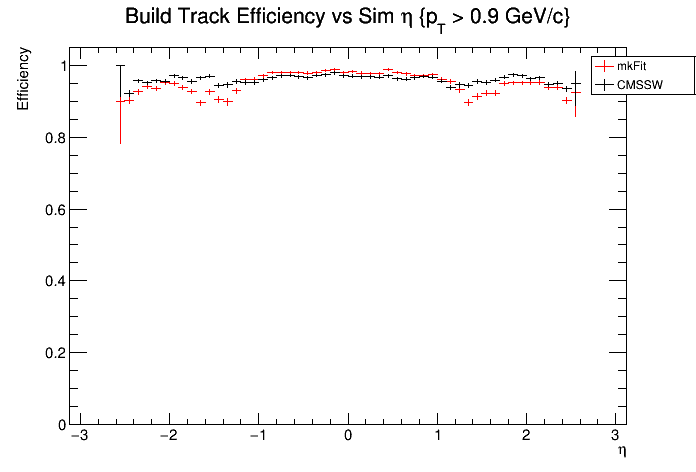}
  \postfigskip

  \caption{Efficiency versus $p_T$ (left) and efficiency versus $\eta$ for
    tracks with $p_t>0.9\GeVoc$, our target $p_T$ limit for CMS HLT operation (right).}
  \label{fig:eff-pt-etapt0p9}
\end{figure}

Track finding efficiency versus $p_T$ and $\eta$ for \mkfit and CMSSW are shown
in figure \ref{fig:eff-pt-etapt0p9}. \mkfit's performance is essentially
equivalent to that of CMSSW for $p_T > 0.9\GeVoc$, the standard CMS cut for detailed
efficiency studies. Below that, \mkfit's
inefficiency is largest in the transition region and noticeable in the
endcaps. We believe that tuning of hit selection windows and candidate scoring
criteria can help us achieve efficiencies comparable to CMSSW for all tracking
regions, down to $p_T = 450\MeVoc$.

\begin{figure}[thb]
  \centering
  \includegraphics[width=\twop]{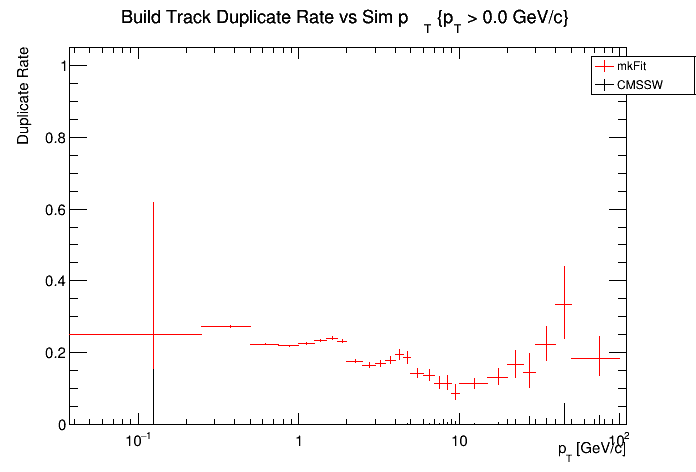}
  \hfill
  \includegraphics[width=\twop]{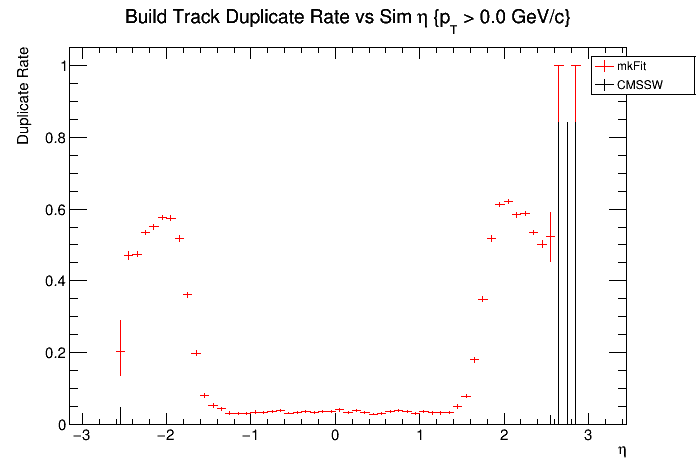}
  \postfigskip

  \caption{Duplicate track rate versus $p_T$ (left) and $\eta$ (right).}
  \label{fig:drates}
\end{figure}

Figure \ref{fig:drates} shows \mkfit's duplicate track rates versus $p_T$ and
$\eta$. CMSSW's duplicate rates are 0. Duplicate rate of \mkfit is significant for all
values of $p_T$. In $\eta$, it is below 5\% level in the barrel and rises
sharply when tracks start entering the endcap disks. The duplicate rate 
distribution is exactly the same as for the artificial 10-muon events, showing that
the duplicate rate is entirely due to duplicate seeds arising from module overlaps in the
seeding layers and the absence of a duplicate track
removal procedure in \mkfit (see section \ref{ssec:cms-event-processing}).

As already mentioned, further work is required to make a more detailed
assessment of \mkfit's performance. However, with \mkfit being available
within the CMSSW, all quality assurance and validation tools developed for CMS
tracking are available for more detailed studies and debugging.


\section{Computational performance}
\label{sec:comp-perf}

Computational benchmarks are shown for our main development platforms:

\begin{itemize}

\item KNL -- Knights Landing -- 64 cores: Intel Xeon Phi CPU 7210 @ 1.30GHz

\item SKL-SP -- Skylake Gold -- 2 sockets x 16 cores: Intel Xeon Gold 6130 CPU @ 2.10GHz

\end{itemize}

While the Turbo Boost feature is turned off on all our development machines,
SKL-SP processor cores feature different frequency characteristics depending
on how many cores are in use and which vector instruction set is being used on
them. KNL can also vary frequency to a lesser extent when AVX-512 code is
being executed.

Results presented in this section were obtained using a subset of the 
events with the configuration described in the introduction to section 
\ref{sec:phys-perf}. The Intel \stt{icc} compiler was used to compile the
code. \mkfit uses Intel Thread Building Blocks (TBB) for multithreading.

\subsection{Single event performance of core track finding}

To assess the performance of the track finding algorithm alone,
we run a dedicated benchmark measuring the track-finding time for
a sufficient number of events ($20 \times \mathrm{N_{threads}}$)
without including the time needed
to pre-process the hits and seeds, or to post-process the track
candidates. This allows us to focus on the most relevant part of our code and
to sideline the more administrative tasks that might, in a production system,
be performed outside of \mkfit itself.

\begin{figure}[htb]
  \centering
  \includegraphics[width=\twop]{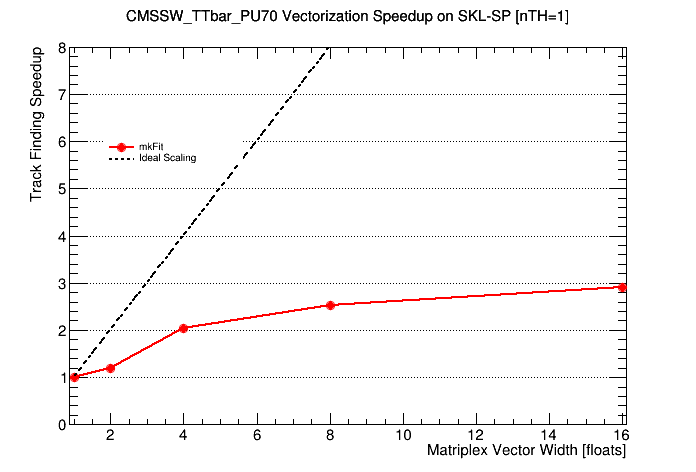}
  \hfill
  \includegraphics[width=\twop]{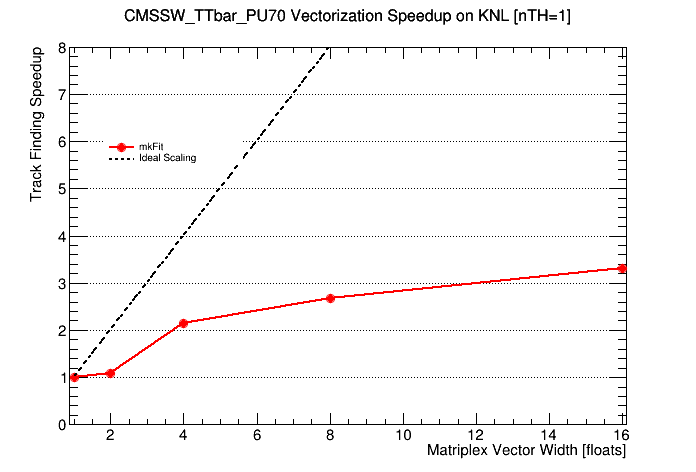}
  \postfigskip

  \caption{Vectorization speedup as a function of used vector width for
    Skylake (left) and Knights Landing (right)
    processors.}
  \label{fig:vu-speedup}
\end{figure}

First, we show the speedup as a function of the Matriplex width which
effectively controls how many slots in the vector registers are used. The
results are shown in figure \ref{fig:vu-speedup}. On both platforms the highest
speedup is obtained with AVX-512 vectorization (SIMD width of 16 floats) and
is 2.9 for SKL-SP and 3.3 for KNL. If we assume that vectorized code obtains
the full speedup given by the Matriplex vector width, and that overall speedup
is impaired only by unvectorized (serial) code sections, then Amdahl's Law
implies that the code must be executing vector instructions at least 72\% of
the time (based on the final speedup of 3 for vector width 16).

\begin{figure}[htb]
  \centering
  \includegraphics[width=\twop]{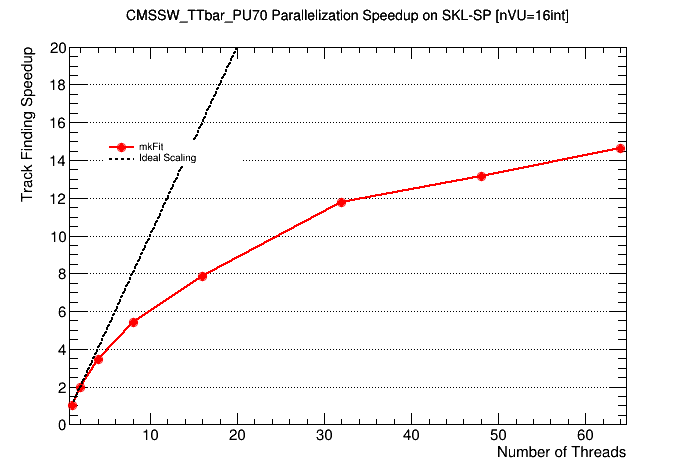}
  \hfill
  \includegraphics[width=\twop]{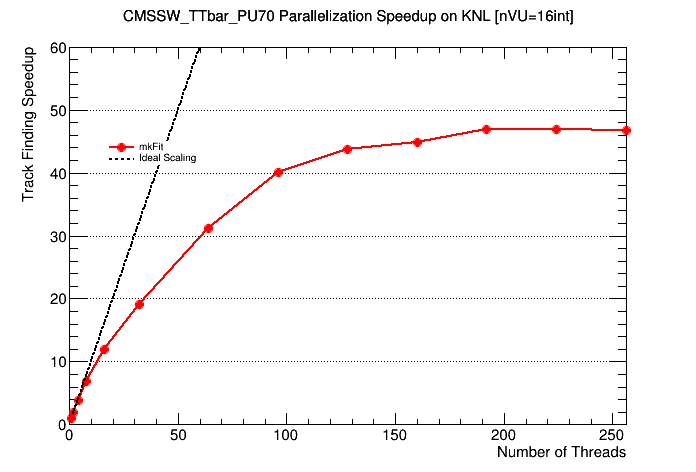}
  \postfigskip

  \caption{Multithreading speedup as a function of used number of threads for
    Skylake (left) and Knights Landing (right) processors.}
  \label{fig:th-speedup}
\end{figure}

Figure \ref{fig:th-speedup} shows speedup as a function of the number of threads
TBB is configured to use. Note that events are processed sequentially and all
parallelism happens during the processing of seeds belonging to the same
event. Beyond 32 threads the standard work chunk of 32 seeds per task gets
progressively reduced down to 4 seeds per task at 256 threads.
See \cite{pkf-tbb} for details about how multithreading is implemented in \mkfit.
SKL-SP shows good scaling up to 8 threads and KNL up to 32 threads.
For KNL the effect of reduction of the standard work chunk can be observed
in continuation of scaling toward higher number of threads, up to 
about 164 threads.

\subsection{Full processing with multiple concurrent events in flight}

To assess the scaling behavior of the full event processing chain as it would
run in CMSSW which can process several events concurrently, we implemented
support for multiple concurrent events in flight in \mkfit as well \cite{pkf-acat-17}.
Technically, this is achieved by using the TBB \stt{parallel\_for} construct
for the event loop itself and retaining all intra-event parallelism described before. This
balances out the tail effects present in event-by-event processing and allows
the tasks themselves to be larger, thus reducing the multithreading overhead. As before, the
number of events processed for each test was 20 times the number of configured threads.

\begin{figure}[htb]
  \centering
  \includegraphics[width=\twop]{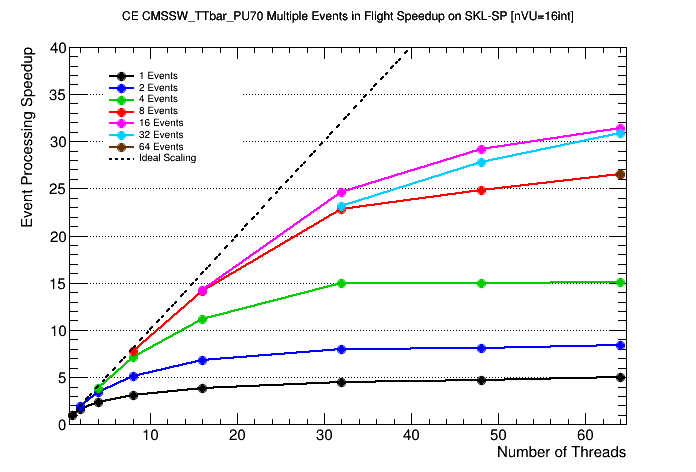}
  \hfill
  \includegraphics[width=\twop]{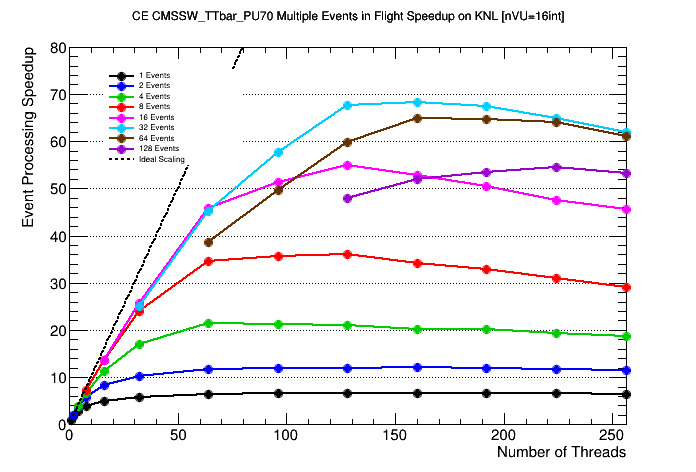}
  \postfigskip

  \caption{Multithreading speedup for different numbers of concurrent events
    in flight, as a function of used number of threads for Skylake (left),
    Knights Landing (right).}
  \label{fig:meif-speedup}
\end{figure}

Scaling behavior for multiple events in flight is shown in figure
\ref{fig:meif-speedup}. Many of the administrative tasks related to
pre-processing of hits and seeds have not yet been fully optimized
or vectorized. One can see the effect of
those by comparing results for one event in flight with corresponding result in
the previous section, figure \ref{fig:th-speedup}. SKL-SP shows best scaling with 16
events in flight; hyperthreading provides additional 30\% speedup when going from 32 to 64
running threads. For KNL, having 32 events in flight offers the best performance; up to
64 threads the same performance is also achieved by 16 events in flight. Having more than
32 events in flight is not helpful, possibly due to the fact that in KNL a given
memory reference can only be ``owned'' by 1 of 32 tiles in the layout of cores.
KNL shows no gain in using more than 128 threads, i.e., hyperthreading does not
yield any additional speedup.

\subsection{Estimated performance of \mkfit at CMS HLT}

In the CMS HLT, due to its processing time constraints, tracking is run only
for a subset of all the input events. On the other hand, running on all events
(100\,kHz rate) a version of the tracking similar to the one used offline
would allow better event selection, cleaner physics data sets and thus better
utilization of storage and CPU resources after data is already
taken. Comparing CMSSW and standalone \mkfit single-threaded performance of
the initial offline tracking iteration, one finds that \mkfit runs about
10-times faster than CMSSW offline tracking. Correspondingly, measured \mkfit
full-node event processing rates for the expected LHC Run 3 pileup of 70 are
115\,events/s for KNL and 250\,events/s for SKL-SP. Thus, to process events at
CMS HLT at the expected 100\,kHz rate, one would need an equivalent of 400
32-core Skylake machines for track reconstruction alone. Note that this is
below the current size of the CMS HLT cluster.

These results are to be considered as very preliminary. In fact, while on one
side we believe the current version of the \mkfit code can be further
optimized, running in the actual HLT configuration within CMSSW requires more
work to mitigate overheads due to data preparation (local reconstruction) and
data conversions.

\section{Conclusion}

Following developments required to support complex, realistic detector
geometries, \mkfit is now in the position to demonstrate its potential for use in
real-world reconstruction scenarios. Preliminary results show that \mkfit
exhibits physics performance on par with existing, traditional KF tracking
algorithms while retaining a significant boost in computational
performance. It also shows the potential to make efficient use of many-core
architectures with few concurrent processes.

Ongoing work is focusing on finishing the tuning of track finding algorithm
parameters and implementing the missing final post-processing of
tracks. Integration with CMSSW is proceeding in parallel with the goal of
integration in the CMS HLT test-bed system for Run 3 of the LHC.

\section{Acknowledgments}

This work is supported by the U.S. National Science Foundation, under the
grants PHY-1520969, PHY-1521042, PHY-1520942 and PHY-1624356,
and by the U.S. Department of Energy, Office of
Science, Office of Advanced Scientific Computing Research, Scientific
Discovery through Advanced Computing (SciDAC) program.


\end{document}